\documentclass[preprintnumbers,article,amsmath,amssymb,floatfix,10pt,prd,onecolumn,
superscriptaddress,nofootinbib]{revtex4}
\usepackage[colorlinks=true, pdfstartview=FitV, linkcolor=blue, citecolor=red, urlcolor=magenta]{hyperref}
\usepackage{bbm}
\usepackage{amsfonts}
\usepackage{mathrsfs}
\usepackage{latexsym}
\usepackage{epsfig}
\usepackage{epstopdf}
\usepackage{epstopdf}
\usepackage{graphicx}
\usepackage{amssymb}
\usepackage{amsmath}
\usepackage{dcolumn}
\usepackage{bm}
\usepackage{float}
\usepackage{color}
\usepackage{comment}
\usepackage{xcolor}
\begin{document}
\title{\bf Viable embedded wormholes and energy conditions in $f(\mathcal{R},\mathcal{G})$ gravity}

\author{Asifa Ashraf} \email{asifa.ashraf70@yahoo.com}\affiliation{ School of Mathematical Sciences, Jiangsu Key Laboratory for NSLSCS, Nanjing Normal
University, Nanjing, 210023, People's Republic of China}

\author{Saadia Mumtaz}
\email{saadia.icet@pu.edu.pk}\affiliation{Institute of Chemical
Engineering and Technology, University of the Punjab, Quaid-e-Azam
Campus, Lahore-54590, Pakistan}

\author{Faisal Javed}
\email{faisaljaved.math@gmail.com}\affiliation{Department of
Physics, Zhejiang Normal University, Jinhua 321004, People's
Republic of China}

\author{Zhiyue Zhang}
\email{zhangzhiyue@njnu.edu.cn}\affiliation{ School of Mathematical Sciences, Jiangsu Key Laboratory for NSLSCS, Nanjing Normal
University, Nanjing, 210023, People's Republic of China}

\begin{abstract}
The current study explores the generalized embedded wormhole solutions in the background of $f(\mathcal{R},\mathcal{G})$ gravity, where $\mathcal{R}$ represents the Ricci scalar and $\mathcal{G}$ denotes the Gauss-Bonnet invariant. To investigate the necessary structures of the wormhole solutions we thoroughly analyzed the energy conditions under $f(\mathcal{R},\mathcal{G})$ gravity within the anisotropic source of matter. To meet this aim, we consider spherically symmetric geometry with the most generic gravity model of the gravity. A modified version of the field equations is calculated for two different embedded wormhole solutions. All the energy conditions are calculated and shown graphically with the regional ranges of the model parameter. Further, the invalid region of the energy conditions confirms the presence of exotic matter. Finally, we have concluding remarks. 
\end{abstract}

\maketitle

\date{\today}

%%%%%%%%%%%%%%%%%%%%%%%%%%%%%%%%%%%%%%%%%%%%%%%%%%%%%%%%%%%%%%%%%%%%%%%%
%%%%%%%%%%%%%%%        Introduction        %%%%%%%%%%%%%%%%%%%%%%%%%%%%%
%%%%%%%%%%%%%%%%%%%%%%%%%%%%%%%%%%%%%%%%%%%%%%%%%%%%%%%%%%%%%%%%%%%%%%%%

\section{Introduction}

The mystery of cosmic expansion is one of the most studied issues
that has brought various revolutions in the modern phase. Many
astronomical probes showed the evidences of accelerated expansion of
universe at its present phase \cite{16a}-\cite{20a}. It has been
suggested that an unusual mysterious source of energy, named as dark
energy (DE), is causing the current accelerated expansion of cosmos.
Researchers have made various attempts to understand the mysterious
nature of DE. Without any favorable candidate for the dark sector,
some alternative ways have been chosen like dynamical DE models,
modified and higher dimensional gravities etc. There are certain
reasons in the literature proposing this phenomenon from the diverse
models of DE to the modified gravity. These approaches can be
classified into two categories, i.e., using matter sources like
Chaplygin gas, quintessence, phantom, quintom, tachyon, k-essence,
cosmological constant etc. or modifying the gravity. Although the
first category is quite interesting but not chosen due to some
uncertainties, while modified theories of gravity can be the best
choice due to their effective cosmological outcomes. Different valid
queries about cosmological constant, initial singularity and
flatness problem can be effectively discussed by modified theories.

Einstein introduced the concept of geometry-matter coupling whose
modifications imply significant outcomes. Taking the original theory
into account, various alternative models have also been presented
through different choices of Lagrangian. The well-known modified
theories of gravity comprise of $f(\mathcal{R})$ gravity \cite{21a}, $f(\mathcal{G})$
gravity \cite{22a}, scalar- tensor theory \cite{23a},  $f(\mathcal{R},T)$
gravity \cite{24a} and $f(\mathcal{R},\mathcal{G})$ gravity \cite{25a}. It is worthwhile
to note that $f(\mathcal{R},\mathcal{G})$ gravity is a remarkable modification of GR
with non-linear combination of scalar curvature $\mathcal{R}$ and Gauss-Bonnet
invariant $\mathcal{G}$. The behavior of $\mathcal{G}$ in curvature invariant
corresponds to the early evolutionary state. Moreover, this theory
describes the accelerating waves of the celestial objects and
evolutionary mechanism from acceleration to deceleration phases.

The study of different cosmic mechanisms illustrates the concept of
hypothetical geometries with topological structures termed as
wormholes. A wormhole (WH) provides a short bridge between any two
regions of the universe thus making a convenient trip to the far
away regions. This theoretical feature has a primal history starting
with Flamm \cite{1a} as a non-traversable WH. Morris and Thorne
\cite{2a} introduced traversable WH configuration that connects
distant regions by a throat. This matter keeps the throat open but
violate the null energy condition that must be reduced for physical
existence of WHs. For physical viability of WH solutions, we must
restrict the amount of exotic matter at the WH throat. There is a
large body of literature for the study of different physical aspects
of WHs constructed from black holes \cite{3a}-\cite{15a}.

The alternative theories of gravity may hold a realistic source for
the WH construction. In these theories, normal fluid fulfill all the
energy constraints whereas the effective stress-energy tensor
corresponds to the violation. Lobo and Oliveira \cite{26a} found the
WH solutions in $f(\mathcal{R})$ gravity by choosing specific
shape-function and equation of the state parameters. Garcia and Lobo
\cite{30a} presented the exact WH solutions in the framework of
Brans-Dicke theory. Azizi \cite{27a} discussed WH solutions in
$f(\mathcal{R},T)$ gravity without violation of the energy
conditions. Static WH solutions were explored in $f(\mathcal{R})$
gravity with the possibility of their existence in the barotropic
matter case \cite{28a}. Sharif and Fatima \cite{31a} found WH
configurations by taking different choices of the shape functions in
$f(\mathcal{G})$ gravity. Zubair et al. \cite{29a} analyzed
spherical WH solutions in $f(\mathcal{R},T)$ gravity by taking
anisotropic fluid distribution. Shamir and his collaborators
\cite{32a,33a} studied some viable WH configurations by choosing
different shape functions and the fluid distribution. Mustafa et al.
\cite{34a} examined non-commutative WH solutions in $f(\mathcal{G},
T)$ gravity. Capozziello et al.
\cite{CZ1,CZ2,CZ3,CZ4,CZ5,CZ6,CZ7,CZ8} explored various  aspects of
WH in alternative gravity theories. Javed et al. \cite{35a} provided
physical analysis of traversable WHs in the context of Rastall
gravity. The stability of WH through thin-shell discussed by Ali and
his coauthors \cite{aa1,aa2,aa3,aa4,aa5} by considering different
black hole solutions.

Inspired by the significant characteristics of $f(\mathcal{R},\mathcal{G})$ formulation,
it is always interesting to discuss some astrophysical issues in
this framework. Thus we are interested to explore the consequences
of this theory in the physical analysis of WH solutions. The paper
is arranged in the following manner. Section II deals with some
basic formalism of the $f(\mathcal{R},\mathcal{G})$ theory and the WH geometry. In
section III, we find new WH solutions in the $f(\mathcal{R},\mathcal{G})$ gravity and
study embedded class of WH solutions by considering the Karmarkar
condition. The last section provides the outcomes.

%%%%%%%%%%%%%%%%%%%%%%%%%%%%%%%%%%%%%%%%%%%%%%%%%%%%%%%%%%%%%%%%%%%%%%%%
%%%%%%%%%%%%%%%            Gravity         %%%%%%%%%%%%%%%%%%%%%%%%%%%%%
%%%%%%%%%%%%%%%%%%%%%%%%%%%%%%%%%%%%%%%%%%%%%%%%%%%%%%%%%%%%%%%%%%%%%%%

\section{$f(\mathcal{R},\mathcal{G})$ Gravity and Field Equations}

An extended version of the action for the current modified theory of gravity like
$f(\mathcal{R},\mathcal{G})$ gravity is provided as \cite{gm4}
\begin{eqnarray}\label{1}
S_{A}=\frac{1}{2K}\int d^4x\sqrt{-g}f(\mathcal{R},\mathcal{G})+S_m(g^{\mu\nu},\psi).\label{1}
\end{eqnarray}
where $g, K$ and $S_m$ represent the detriment of metric function
$g^{\mu\nu}$, coupling constant and the matter source, respectively
while $f(\mathcal{R},\mathcal{G})$ is the function of Gauss-Bonnet
invariant and Ricci scalar. By making an appropriate variation
\cite{17} of the action by Eq. (\ref{1}), the respective field
equations in $f(\mathcal{R},\mathcal{G})$ gravity  are given as
\begin{eqnarray}
\mathcal{R}_{\mu\nu}-\frac{1}{2}g_{\mu\nu}\mathcal{R}&=&K T^{(m)}_{\mu\nu}
+\bigg(\nabla_\mu\nabla_\nu f_\mathcal{R}-g_{\mu\nu}\Box f_\mathcal{R}
+\mathcal{R}\nabla_\mu\nabla_\nu f_\mathcal{G}+4\mathcal{R}_{\mu\nu}\Box
f_\mathcal{G}+4g_{\mu\nu}\mathcal{R}^{\theta\phi}\nabla_\theta\nabla_\phi
f_\mathcal{G}-4\mathcal{R}^\alpha_\nu\nonumber
\\&\times& \nabla_\alpha\nabla_\mu f_\mathcal{G}-2g_{\mu\nu}\mathcal{R}
\Box f_\mathcal{G} -4\mathcal{R}^\alpha_\mu\nabla_\alpha\nabla_\nu
f_\mathcal{G}+4\mathcal{R}_{\mu\theta\phi\nu}\nabla^\theta\nabla^\phi
f_\mathcal{G}-\frac{1}{2}g_{\mu\nu}X+(1-f_\mathcal{R})\mathcal{G}_{\mu\nu}\bigg),
\label{2}
\end{eqnarray}
where
\begin{equation*}
f_\mathcal{R}=\frac{\partial f}{\partial \mathcal{R}},\;\;\;\;\;\;f_\mathcal{G}
=\frac{\partial f}{\partial \mathcal{G}},\;\;\;\;\;X= f_{\mathcal{R}}\mathcal{R}
+f_{\mathcal{G}}\mathcal{G}-f,
\end{equation*}
and $T_{\mu\nu}^{(m)}$ is used as source of the ordinary matter. We
consider a spherically symmetric spacetime for the WH structure
given by
\begin{equation}\label{3}
 d{s}^2=-e^{\epsilon(r)}dt^2+e^{\varepsilon(r)}d{r}^2+r^2(d\theta^2+\sin^2 \theta
 d\varphi^2).
\end{equation}
Here
\begin{itemize}
\item $\epsilon(r)=2\varphi(r)$ with $\varphi(r)$ representing the red-shift function.
\item $e^{\varepsilon(r)}=\left(\frac{r+b(r)}{r}\right)^{-1}$, with $b(r)$ as
the shape function.
\item The WH throat connects two asymptotically flat regions at
$r_0$ (radial coordinate), where $b(r_0)= r_0$.
\item The shape function $b(r)$ should execute the flare-out condition
$\frac{b(r) - rb'(r)}{2b^2(r)} > 0$. This turns to be $b'(r_0)<1$ at
or near the WH throat.
\item The shape function must also fulfill $1-\frac{b(r)}{r}>0$
with $r>r_0$ for sake of maintenance of the metric's signature.
\item For asymptotically flat geometries, we must have the metric functions
obeying this condition, i.e., $\varphi(r)$ and $b(r)/r$ disappear
(becomes zero) as $r$ reaches $\infty$. This criteria can obviously
be neglected in case of non-asymptotically flat geometry.
\end{itemize}
For the current analysis, we assume an anisotropic matter
distribution as follows
\begin{equation}\label{4}
T_{\mu}^{\nu}=\left(\rho+p_t\right)u_{\mu}\,u^{\nu}-p_t\,\delta_{\mu}^{\nu}
+\left(p_r-p_t\right)v_{\mu}\,v^{\nu},
\end{equation}
where $\rho$, $u_{\mu}$, $v_{\mu}$, $p_r$ and $p_t$ denote the
energy-density, four-velocity, the unitary space-like vector, radial
and tangential pressures, respectively. In the present analysis, we
take a model for $f(\mathcal{R},\mathcal{G})$ gravity given by
\cite{gm4}
\begin{equation}\label{5}
f(\mathcal{R},\mathcal{G})=\mathcal{R}+\lambda \times \mathcal{R}^{2}+\mathcal{G}^2,
\end{equation}
where $\lambda$ is a model parameter. The Ricci scalar and
Gauss-Bonnet invariant are calculated as
\begin{eqnarray*}
\mathcal{R}&=&\left[\frac{1}{2} e^{-\varepsilon(r)} \left(2 \epsilon''(r)
-\epsilon'(r) \varepsilon'(r)+\epsilon'(r)^2+\frac{4 \epsilon'(r)}{r}-\frac{4
\varepsilon'(r)}{r}-\frac{4 e^{\varepsilon(r)}}{r^2}+\frac{4}{r^2}\right)\right],\\
\mathcal{G}&=&\left[\frac{2 e^{-2 \varepsilon(r)}
\left(\left(e^{\varepsilon(r)}-3\right)\epsilon'(r)
\varepsilon'(r)+\left(1-e^{\varepsilon(r)}\right) \left(2
\epsilon''(r)+\epsilon'(r)^2\right)\right)}{r^2}\right],
\end{eqnarray*}
In this study, we assume the
specific form of the redshift function to avoiding the any difficulty, which is defined as
\begin{equation}\label{red}
\varphi=-\frac{\zeta }{r},
\end{equation}
where $\chi$ is constant. By plugging Eqs.(\ref{3}-\ref{5}) and Eq.
(\ref{red}) in Eq. (\ref{2}), the respective modified filed
equations for $f(\mathcal{R},\mathcal{G})$ gravity within the WH
geometry are given as
\begin{eqnarray}
\rho&=&\frac{1}{2 r^{16}}\bigg(-2 r^4 b(r) \left(16 \zeta  r (79 r-8 \zeta )
b'(r)^3+\psi _5 b'(r)-32 \zeta  r b'(r)^2 \left(3 r^2 b''(r)+21 \zeta -161 r\right)
+r^2 \psi _6\right)\\&+&\zeta  r^2 \psi _2 b(r)^2-32 \zeta r\psi _1 b(r)^3
+32\zeta \psi _4 b(r)^4+r^6 \psi _3\bigg),\label{13}\\
p_{r}&=&\frac{1}{2 r^{16}}\bigg(-2 r^4 \psi _9 b(r)+\zeta  r^2 \psi _8 b(r)^2
+32 \zeta  b(r)^4 \left(-4 \zeta ^3+98 r^3+409 \zeta  r^2-132 \zeta ^2 r\right)
+64 \zeta  r \psi _7 b(r)^3+r^5 \psi _{10}\bigg),\label{14}\\
p_{t}&=&\frac{1}{2 r^{17}}\bigg(r^4 \left(-\left(24 \zeta  r^3 (2
\zeta +r)b'(r)^4-16 \zeta  r^2 \left(14 \zeta ^2+6 r^3+(6 \zeta +4)
r^2-21 \zeta  r\right) b'(r)^3-r \psi _{15}
b'(r)^2\right.\right.\\&+&\left.\left.\psi _{16} b'(r)+2 r^2 \psi
_{17}\right)\right)+r^3 \psi _{14} b(r)+\zeta  r^2 \psi _{13}
b(r)^2+8 \zeta  r \psi _{12} b(r)^3+8 \zeta  \psi _{11}
b(r)^4\bigg).\label{16}
\end{eqnarray}
where $\psi _j$, $(j=1,...,17)$ are provided in the Appendix (\textbf{I}).

\section{Energy Conditions} \label{sec3}

The energy conditions play an important role like a necessary tools
to check the nature of the matter in the formwork of modifiable
proposals of gravity, especially $f(\mathcal{R},\mathcal{G})$
gravity. Further all the energy conditions are more helpful to
analyze the nature and the geodesic structure of spherically
symmetric space-time. In the literature, there are five kind of
energy conditions exist in the background of modified theories of
gravity and in GR also. The energy bounds are defined as
\begin{eqnarray}
\nonumber  \text{Strong energy condition (SEC)}&\Leftrightarrow&(T_{\epsilon\varepsilon}+\frac{T}{2}g_{\epsilon\varepsilon})X^\epsilon X^\varepsilon\geq 0,~~DEC\Leftrightarrow T_{\epsilon\varepsilon}X^\epsilon X^\varepsilon\geq 0,\\
\nonumber \text{Null energy condition (NEC)}&\Leftrightarrow&T_{\epsilon\varepsilon}\chi^\epsilon \chi^\varepsilon\geq 0,~~~~~~~~~~~~~~~WEC\Leftrightarrow T_{\epsilon\varepsilon}X^\epsilon X^\varepsilon\geq 0,\\
\nonumber \text{Trace energy condition
(TEC)}&\Leftrightarrow&(T_{\epsilon\varepsilon}-\frac{T}{2}g_{\epsilon\varepsilon})X^\epsilon
X^\varepsilon\geq 0,
\end{eqnarray}
where $\chi^\epsilon$ is the null vector and $X^\epsilon$ is a time-like vector. For $DEC$, $T_{\epsilon\varepsilon}X^\epsilon$ is not space like.

For principal pressure, we have
\begin{eqnarray}
\nonumber SEC&&\Leftrightarrow \forall j,~\rho+p_j\geq 0,~\rho+\sum_jp_j\geq 0,\\
\nonumber DEC&& \Leftrightarrow\rho\geq 0,~~\forall j,~p_j\epsilon[-\rho,+\rho],\\
\nonumber NEC&&\Leftrightarrow\forall j,~\rho+p_j\geq 0,~WEC\Leftrightarrow\rho\geq 0,~~\forall j,~\rho+p_j\geq 0,\\
\nonumber TEC&&\Leftrightarrow \forall j,~\rho-p_j\geq
0,~\rho-\sum_jp_j\geq 0,
\end{eqnarray}
which yield
\begin{eqnarray}
\nonumber SEC&:&\rho+p_{r}\geq 0,~~~~~~\rho+p_t\geq 0,~~~~~~~\rho+p_r+2p_t\geq 0,\\
\nonumber DEC&:&\rho\geq 0,~~~~~~~~~~~~\rho-|p_r|\geq 0,~~~~~~\rho-|p_t|\geq 0,\\
\nonumber NEC&:&\rho+p_r\geq 0,~~~~~\rho+p_t\geq 0,\\
\nonumber WEC&:&\rho\geq 0,~~~~~~~~~~~~\rho+p_r\geq 0,~~~~~~~\rho+p_t\geq 0,\\
\nonumber TEC&:&\rho-p_{r}\geq 0,~~~~~~\rho-p_t\geq 0,~~~~~~~\rho-p_r-2p_t\geq 0.
\end{eqnarray}

The energy constraints are verified for normal matter distributions.\\

\section{Embedded Wormhole Solutions}

In this paper, we look into embedded WH models using two different
methodologies. Firstly, we shall start with the comprehensive
embedded class of WH solutions by considering the Karmarkar
condition \cite{Karmarkar/1948} under class-1 and Ellis-Bronniokv
embedded space-time. The fundamental structure of the Karmarkar
condition be contingent on the class-1 embedded solution of
Riemannian space. Eisenhart inspected the suitable condition for the
class-1 embedded solution \cite{Eisenhart/1966}, which comes with
Gauss equation defined as
\begin{eqnarray}\label{r10}
\mathcal{R}_{mnpq}=2\,\epsilon\,{b_{m\,[p}}{b_{q]n}}.
\end{eqnarray}
The Codazzi equation yields
\begin{eqnarray}\label{r11}
b_{m\left[n;p\right]}={\Gamma}^q_{\left[n\,p\right]}\,b_{mq}
-{{\Gamma}^q_{m}}\,{}_{[n}\,b_{p]q}.
\end{eqnarray}
In the above equation, square brackets are used for
anti-symmetrization, $\epsilon=\pm1$, and $b_{mn}$ are the
coefficients. The Karmarkar condition, through Eqs.(\ref{r10}) and
(\ref{r11}), takes the form
\begin{equation}\label{r12}
R_{2323}R_{1414}=R_{1224}R_{1334}+ R_{1212}R_{3434},
\end{equation}
where $R_{2323}\neq R_{1414}\neq0$. By taking the suitable Riemanian
tensor in Eq.(\ref{r12}), we have
\begin{equation}\label{r13}
\left\{2 \left[\lambda''(r)+\lambda'(r)^2\right]-\lambda'(r)
\nu'(r)-\lambda'(r)^2 \right\}+\frac{\lambda'(r)
\nu'(r)}{1-e^{\nu(r)}}=0,\;\;\;\;\;\;\;e^{\nu(r)}\neq1,
\end{equation}
Solving Eq.(\ref{r13}), we get
\begin{equation}\label{r14}
e^{\nu(r)}=1+\Psi e^{\lambda(r)}\lambda^{'2}(r),
\end{equation}
where $\Psi$ is an integration constant. Now, by adopting the
process reported in \cite{gm1}, the embedded shape function yields
\begin{equation}\label{r15}
b(r)=r-\frac{r^{5}}{r^{4}+b_{0}^{4}(b_{0}-\omega)}+\omega,\;\; 0<\omega<b_{0}.
\end{equation}
where $b_{0}$ is WH throat radius. Now we discuss about the
generalized Ellis-Bronnikov space-time \cite{gm2,gm3} (ultra-static WH
model), which is the second embedded WH solution defined as
\begin{equation}\label{r16}
d s^{2}=-d t^{2}+d l^{2}+r^{2}(l)\left[d \theta^{2}+\sin ^{2}(\theta) d \phi^{2}\right],
\end{equation}
with
\begin{equation}\label{r17}
r(l)=\left[b_{0}^{m}+l^{m}\right]^{1 / m}.
\end{equation}
In the above equations, $l$ represents the proper radial distance or
known as tortoise coordinate and can be used for both embedded
solutions. As already mentioned for the previous case that $X_{0}$
is the throat radius of the embedded WH and $m$ is used for the WH
parameter with condition $(m \geq 2)$. Now, Eq.(\ref{5}) can be
rewritten as
\begin{equation}\label{r18}
d s^{2}=-d t^{2}+\frac{d r^{2}}{\left[1-\frac{b(r)}{r}\right]}+r^{2}\left(d
\theta^{2}+\sin ^{2} \theta d \phi^{2}\right].
\end{equation}
The radial coordinate $r$ and radial distance $l$ can be related by
the following embedding relation
\begin{equation}\label{r19}
d l^{2}=\frac{d r^{2}}{\left[1-\frac{b(r)}{r}\right]}.
\end{equation}
Finally, we get the embedded shape function in the form
\begin{equation}\label{r20}
b(r)=r-r^{(3-2 m)}\left(r^{m}-b_{0}^{m}\right)^{(2-\frac{2}{m})}.
\end{equation}
For $m=2$, one gets Ellis-Bronniokv WH geometry with horizonless
space-time. To complete the investigation, we use two distinct
embedded WH solutions with Eqs.(\ref{r15}) and (\ref{r20}) by
merging the radial coordinate $r$ and radial distance $l$ through
Eq.(\ref{r17}).
\begin{figure}[H]
\centering \epsfig{file=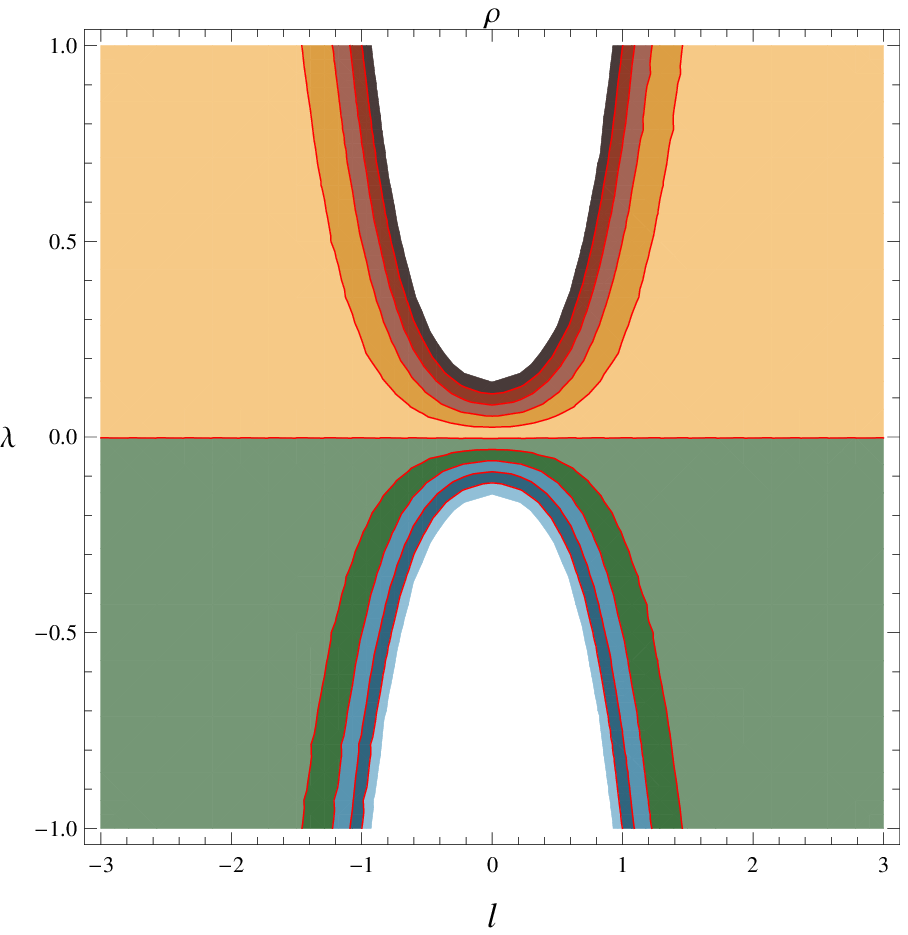, width=.4\linewidth,
height=2.1in}\epsfig{file=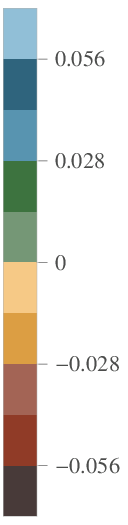, width=.08\linewidth,
height=2.1in}\epsfig{file=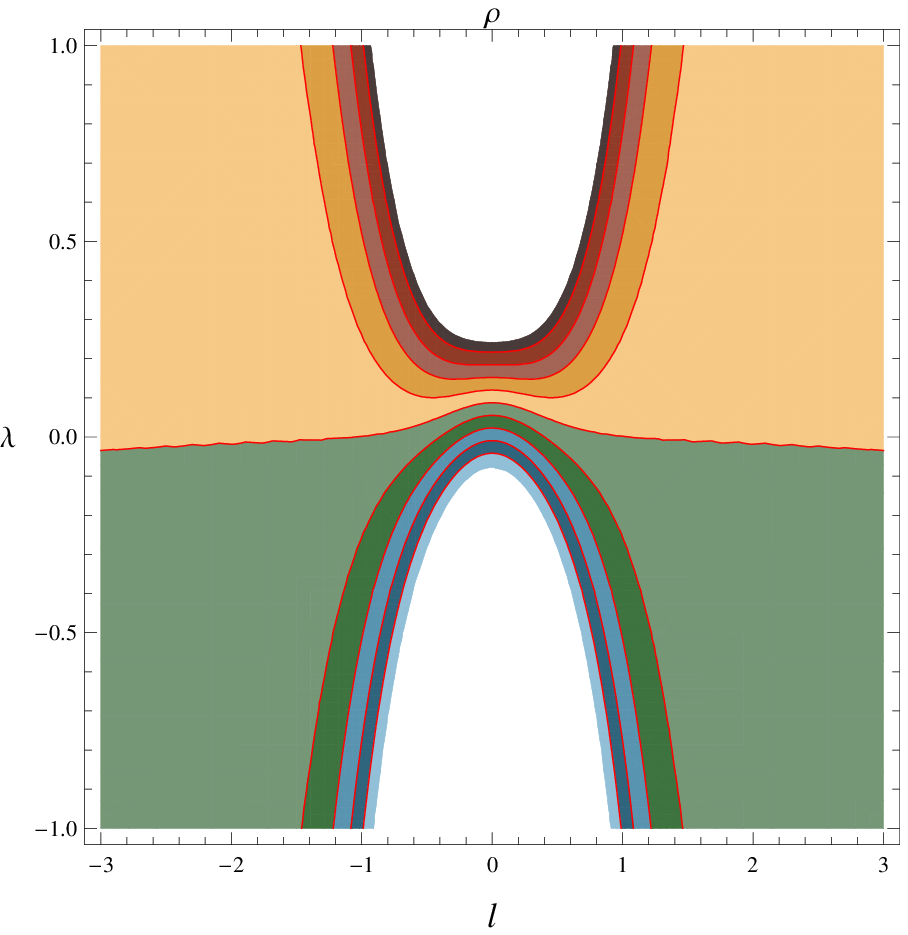, width=.4\linewidth,
height=2.1in}\epsfig{file=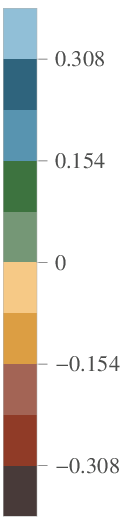, width=.08\linewidth,
height=2.1in} \caption{\label{F1} $\rho$ for embedded shape function
-I (left) and embedded shape function-II (right).}
\end{figure}

\begin{figure}[H]
\centering \epsfig{file=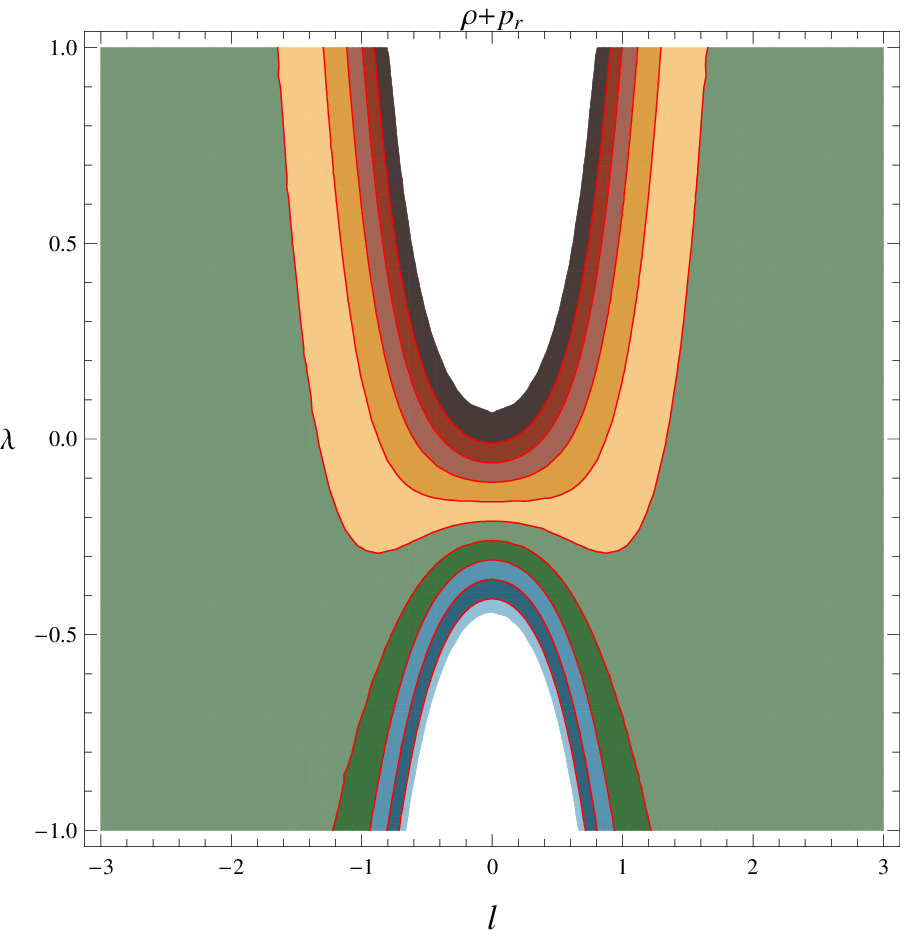, width=.4\linewidth,
height=2.1in}\epsfig{file=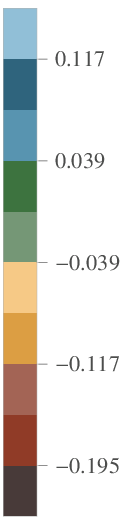, width=.08\linewidth,
height=2.1in}\epsfig{file=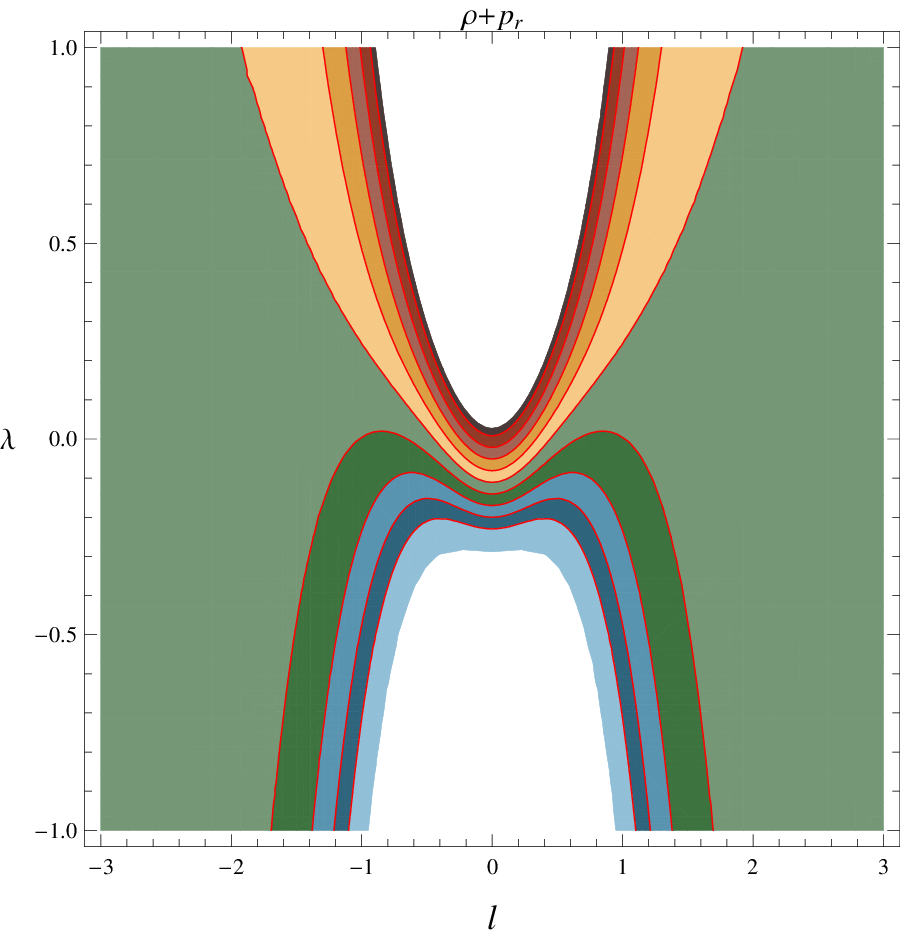, width=.4\linewidth,
height=2.1in}\epsfig{file=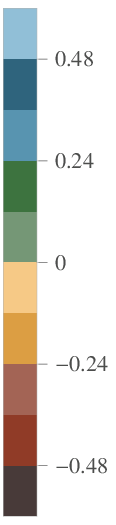, width=.08\linewidth,
height=2.1in} \caption{\label{F2} $\rho+p_{r}$ for embedded shape
function -I (left) and embedded shape function-II (right).}
\end{figure}

\begin{figure}[H]
\centering \epsfig{file=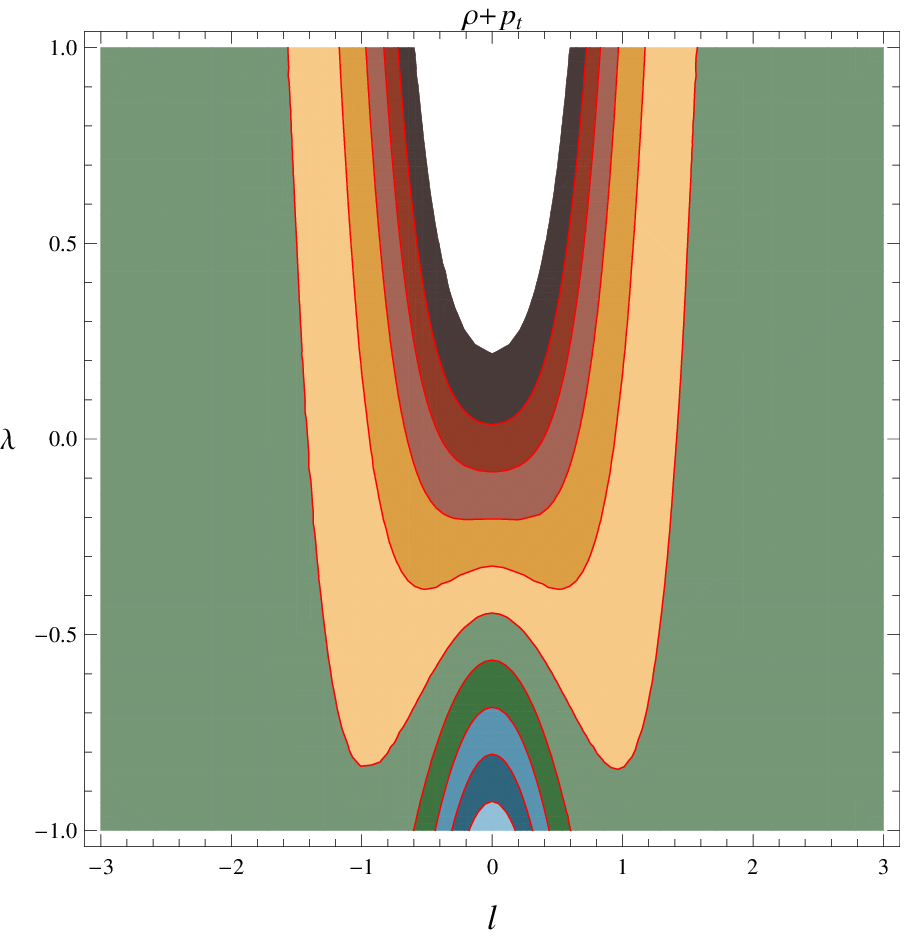, width=.4\linewidth,
height=2.1in}\epsfig{file=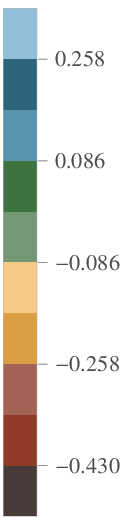, width=.08\linewidth,
height=2.1in}\epsfig{file=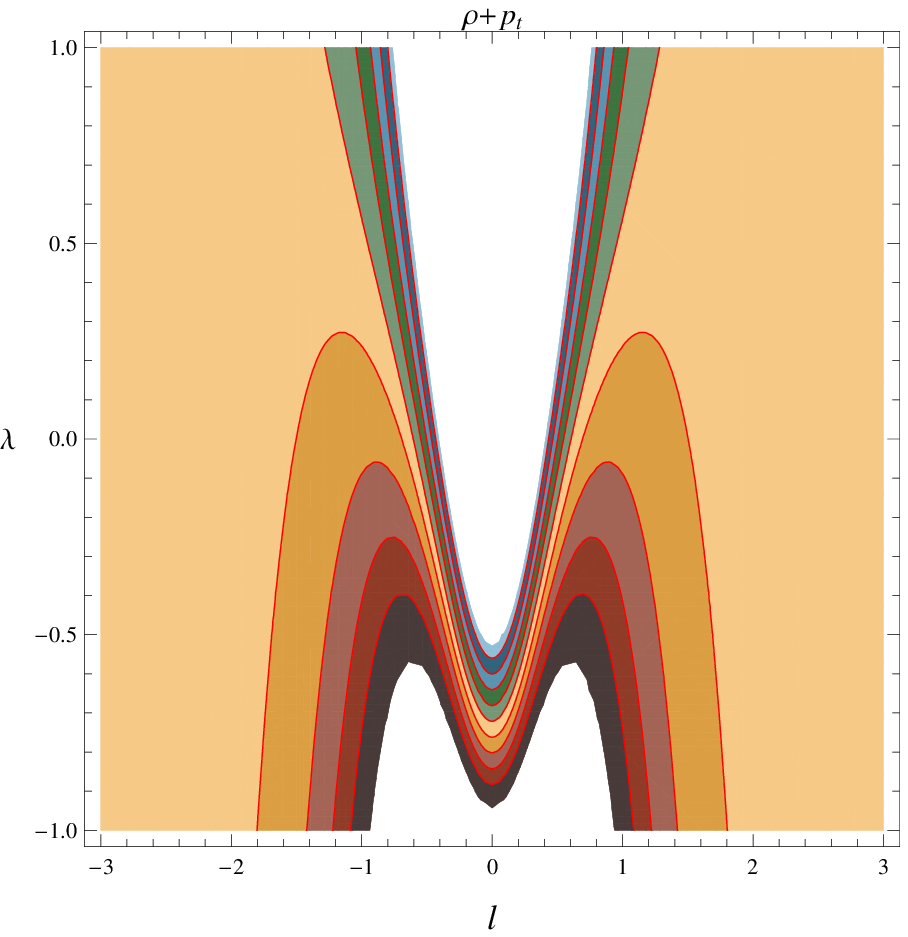, width=.4\linewidth,
height=2.1in}\epsfig{file=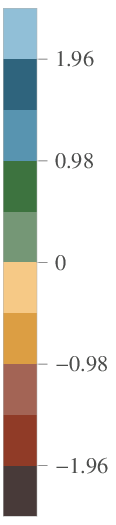, width=.08\linewidth,
height=2.1in} \caption{\label{F3} $\rho+p_{t}$ for embedded shape
function -I (left) and embedded shape function-II (right).}
\end{figure}

\begin{figure}[H]
\centering \epsfig{file=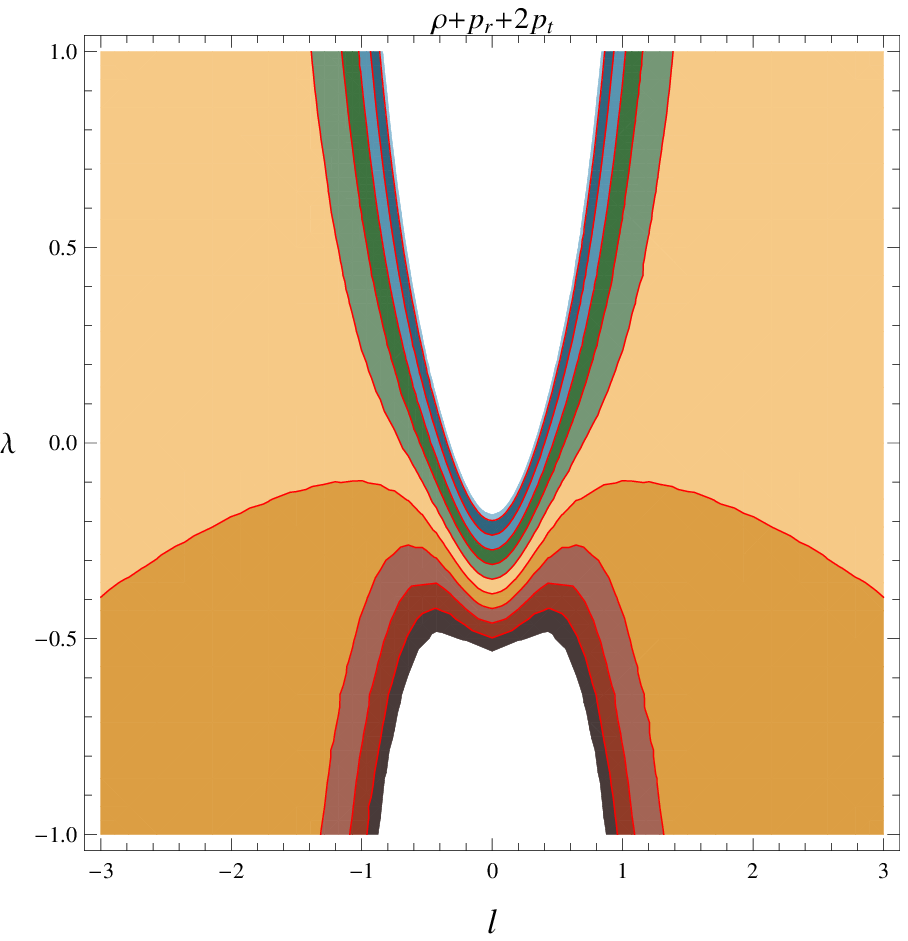, width=.4\linewidth,
height=2.1in}\epsfig{file=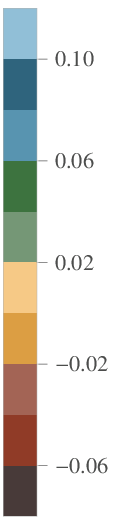, width=.08\linewidth,
height=2.1in}\epsfig{file=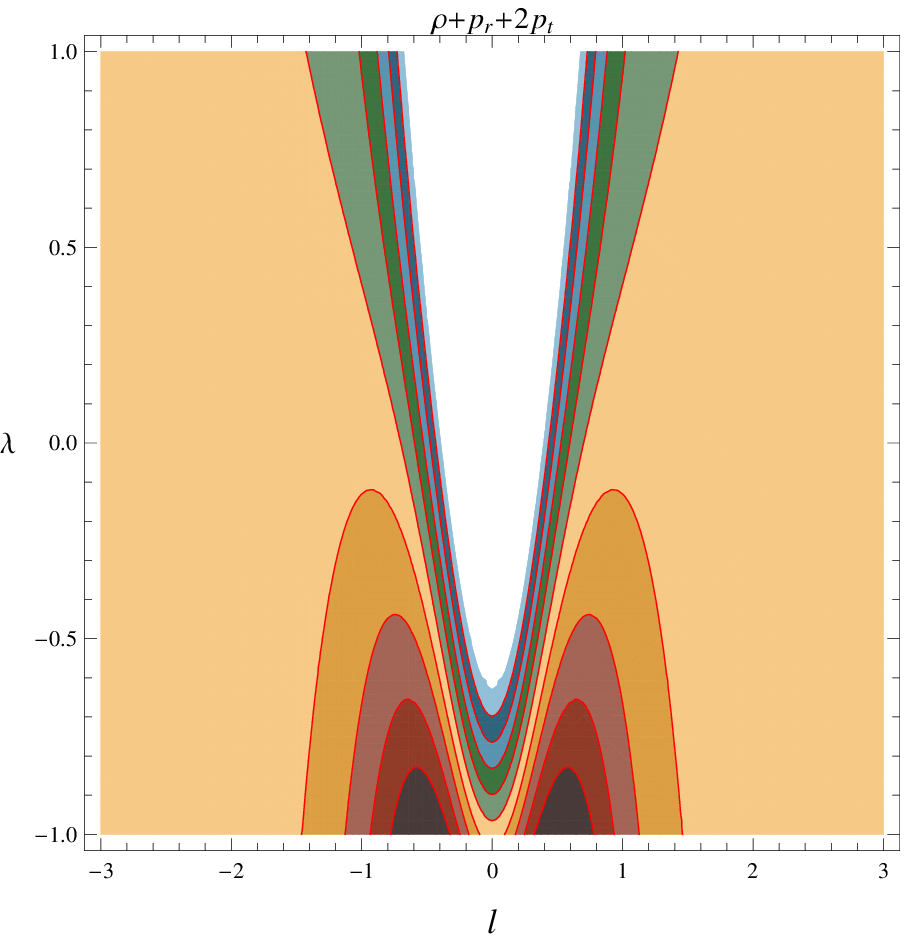, width=.4\linewidth,
height=2.1in}\epsfig{file=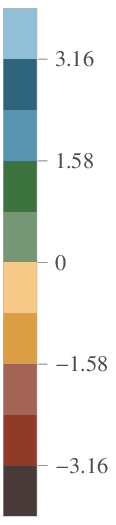, width=.08\linewidth,
height=2.1in} \caption{\label{F4} $\rho+p_{r}+2p_{t}$ for embedded
shape function -I (left) and embedded shape function-II (right).}
\end{figure}

\begin{figure}[H]
\centering \epsfig{file=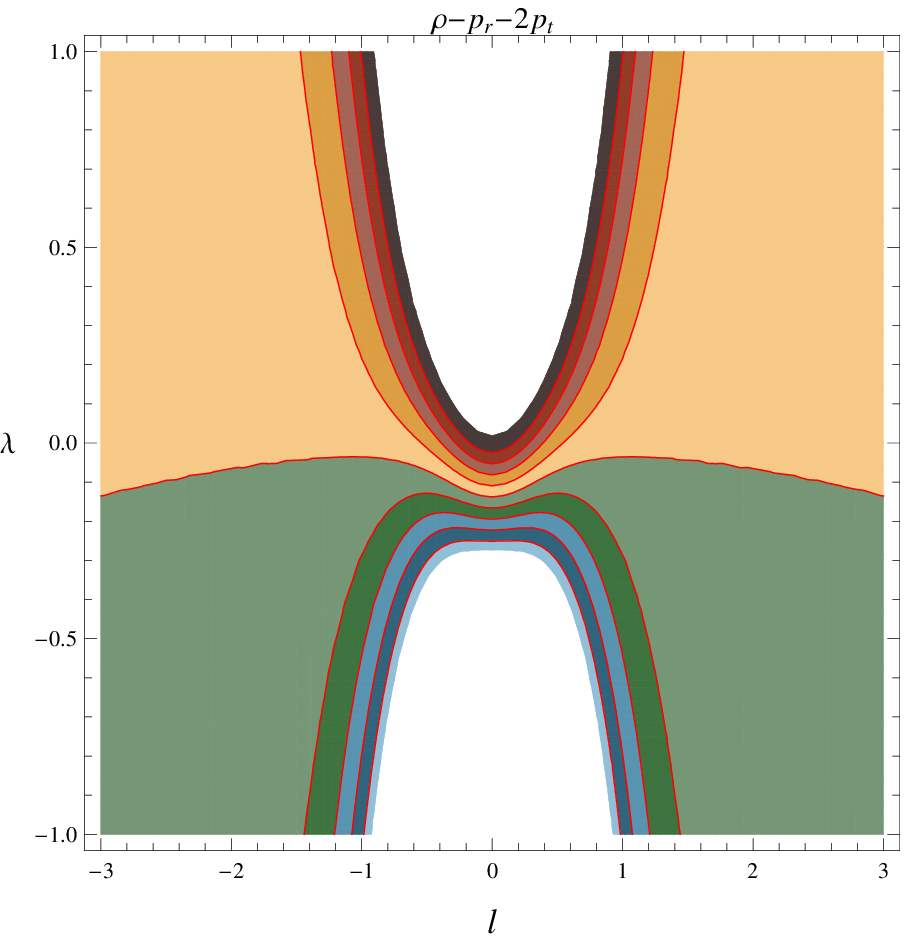, width=.4\linewidth,
height=2.1in}\epsfig{file=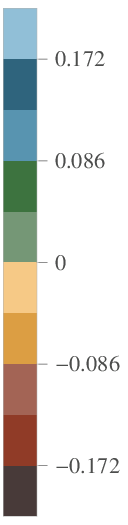, width=.08\linewidth,
height=2.1in}\epsfig{file=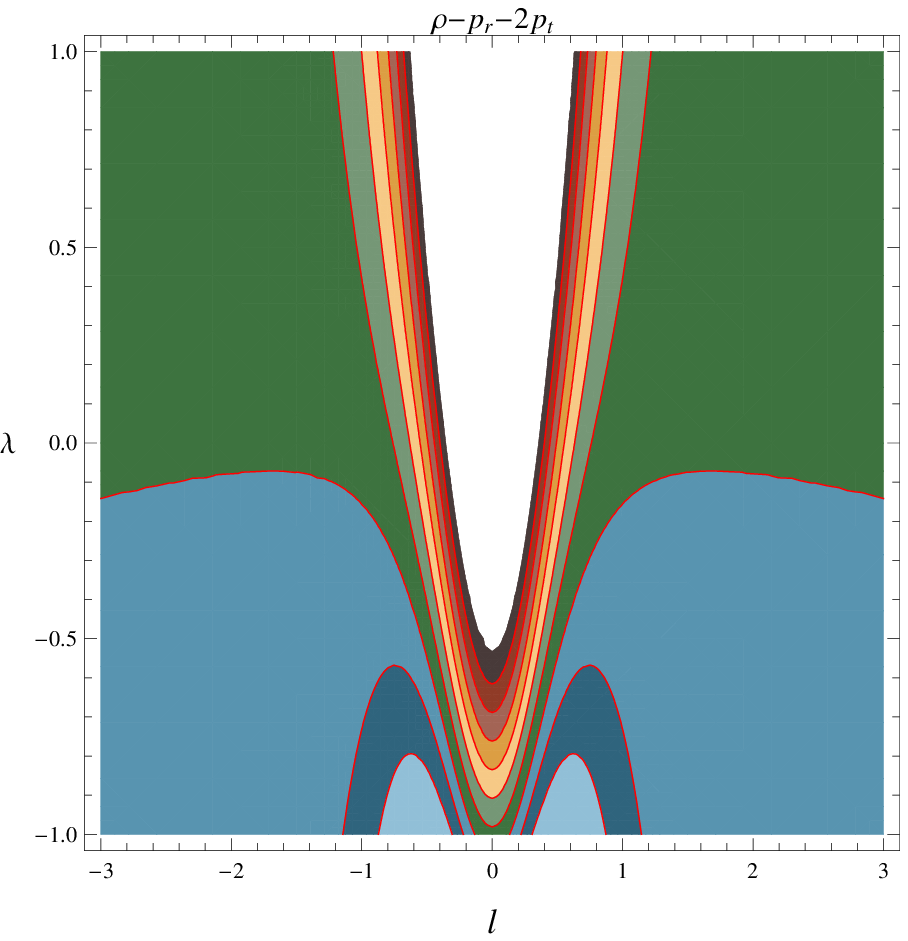, width=.4\linewidth,
height=2.1in}\epsfig{file=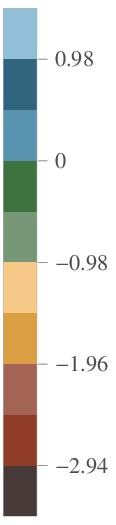, width=.08\linewidth,
height=2.1in} \caption{\label{F5} $\rho-p_{r}-2p_{t}$ for embedded
shape function -I (left) and embedded shape function-II (right).}
\end{figure}

\begin{figure}[H]
\centering \epsfig{file=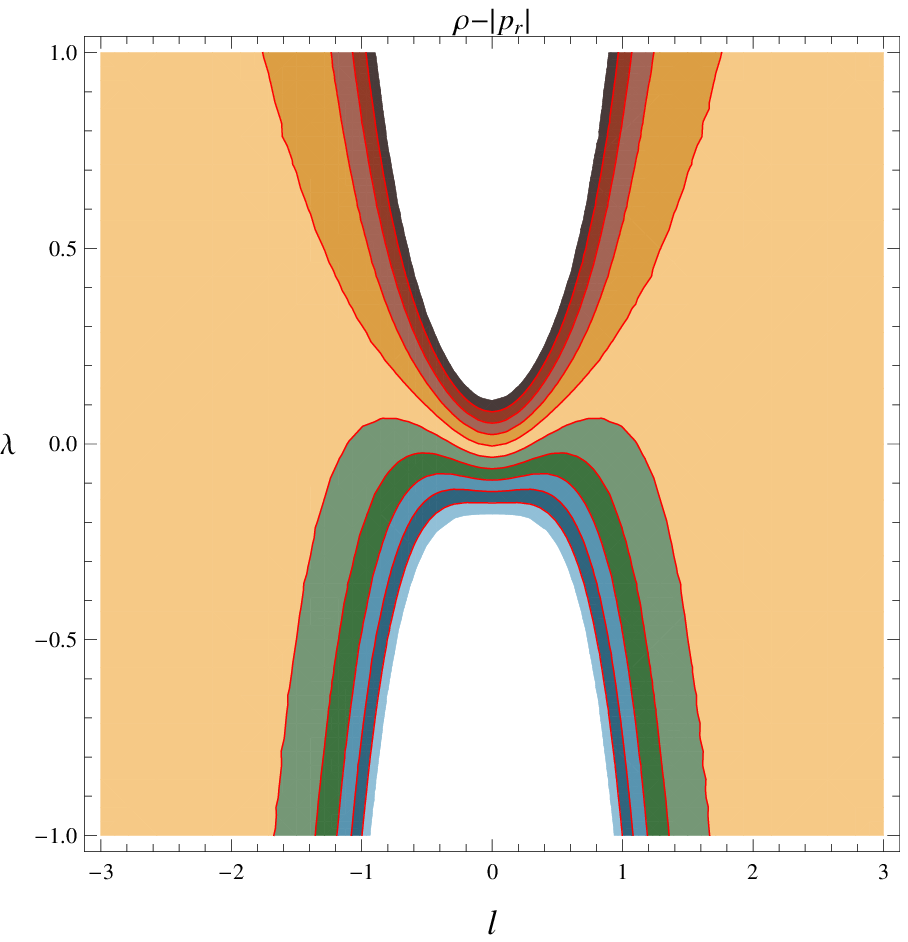, width=.4\linewidth,
height=2.1in}\epsfig{file=Fig5a.eps, width=.08\linewidth,
height=2.1in}\epsfig{file=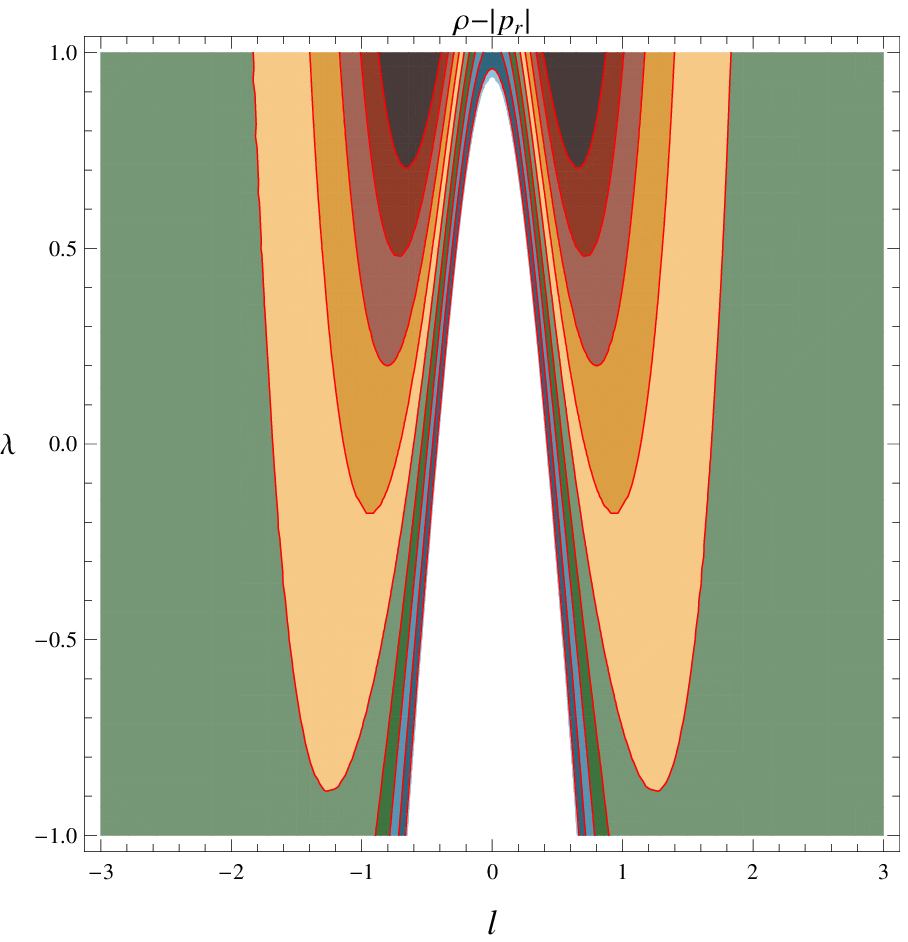, width=.4\linewidth,
height=2.1in}\epsfig{file=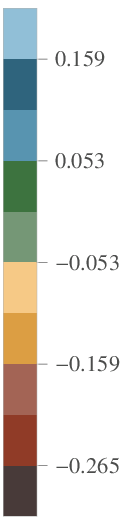, width=.08\linewidth,
height=2.1in} \caption{\label{F6} $\rho-\mid p_{r}\mid$ for embedded
shape function -I (left) and embedded shape function-II (right).}
\end{figure}

\begin{figure}[H]
\centering \epsfig{file=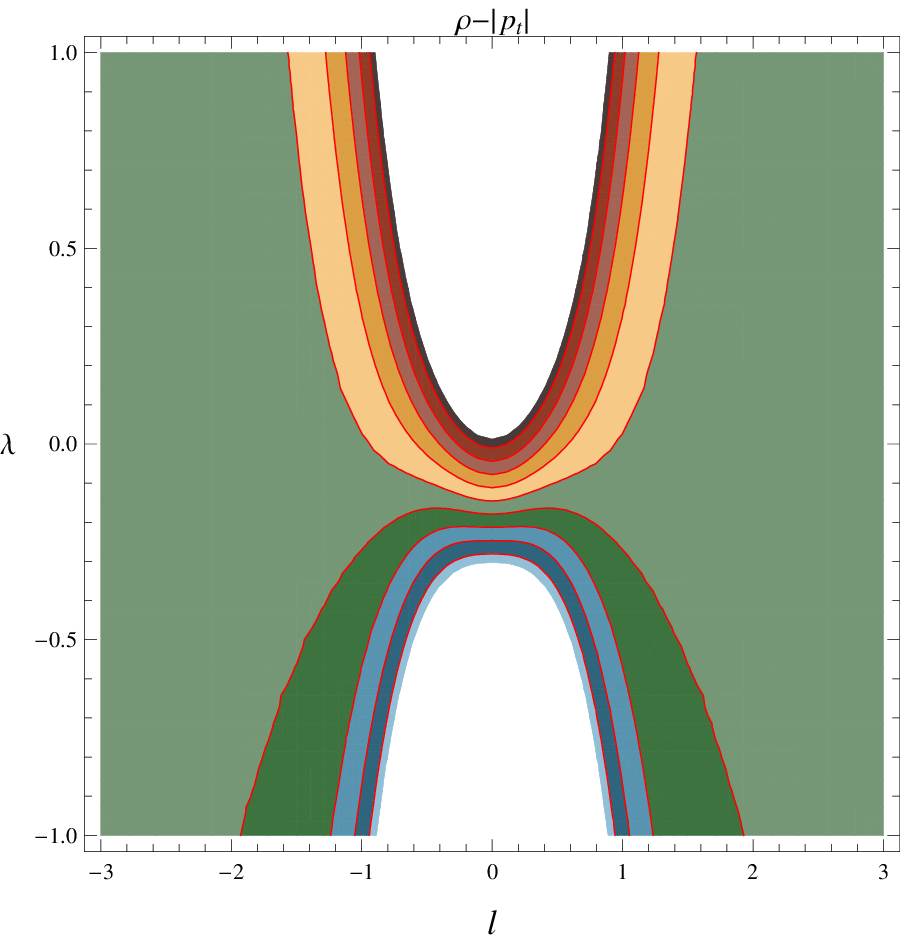, width=.4\linewidth,
height=2.1in}\epsfig{file=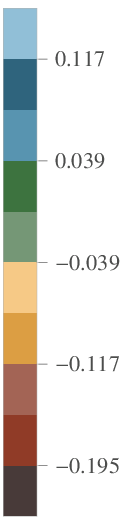, width=.08\linewidth,
height=2.1in}\epsfig{file=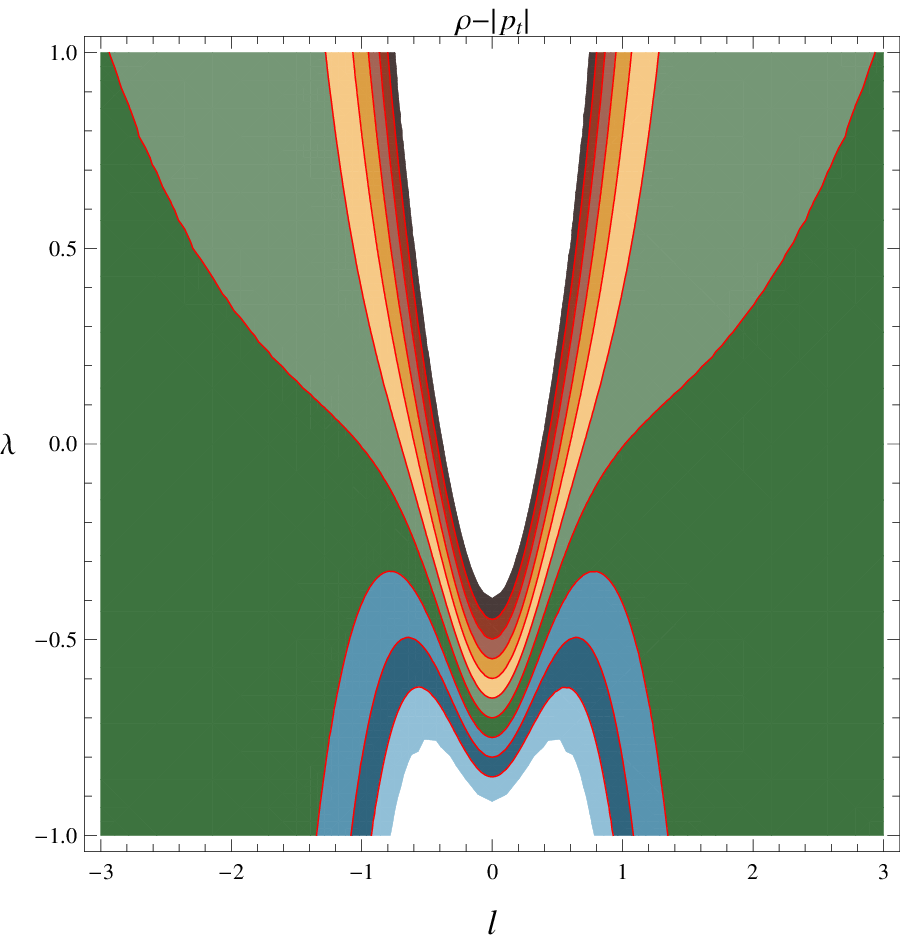, width=.4\linewidth,
height=2.1in}\epsfig{file=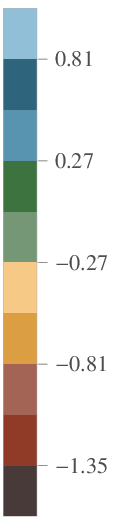, width=.08\linewidth,
height=2.1in} \caption{\label{F7} $\rho-\mid p_{t}\mid$ for embedded
shape function -I (left) and embedded shape function-II (right).}
\end{figure}
In order to explore the current analysis, we try to provide the
regional graphical analysis to check the possibility of the WH
existence thorough the nature of the matter. It is necessary to
mention that we are only exploring the nature of energy conditions
in $f(\mathcal{R},\mathcal{G})$ for two different embedded solutions
of WH geometry by considering the specific model of the red-shift.
It is also necessary to write about the model parameter like
$\lambda$ that plays an astonishing role in the present study. The
different values of the $\lambda$ provide us the different
possibilities of the WH existence. In the current analysis, we are
going to discuss the all the possibilities of the results in the
specific range of the involved parameter $\lambda$ like $-1\leq
\lambda\leq 1$. In the described range of the model parameter, we
have found the different possibilities of the WH existence. From the
Fig. (\ref{F1}), the behavior of the energy density can be checked
for $-1\leq \lambda\leq 1$ with the radial distance $-\leq l\leq1$
through Eq.(\ref{r17}). The null energy condition, i.e.,
$\rho+p_{r}$ is presented graphically in Fig. (\ref{F2}). It is
noticed from the Fig. (\ref{F2}) that is strongly violated with
$-1\leq \lambda\leq 1$ for radial distance $-\leq l\leq1$ for
embedded shape function-I (left) and embedded shape function-II
(right). Another important energy condition like $\rho+p_{t}$ with
graphically development is provided in Fig. (\ref{F3}). The
condition $\rho+p_{t}$ is also violated in the small ranges of
parameter, $\lambda$. The most important energy conditions like a
strong energy conditions $\rho+p_{r}+2p_{t}$ and trace energy
condition $\rho-p_{r}-2p_{t}$ are given in Fig. (\ref{F4}) and Fig.
(\ref{F5}), respectively. Both the strong and trace energy
conditions are seen violated and confirm the possibility of WH
existence for embedded shape function -I (left) and embedded shape
function-II (right). Some other energy conditions like $\rho-\mid
p_{r}\mid$ and $\rho-\mid p_{t}\mid$ are also presented graphically
in Fig. (\ref{F6}) and Fig. (\ref{F7}), respectively.
\begin{figure}[H]
\centering \epsfig{file=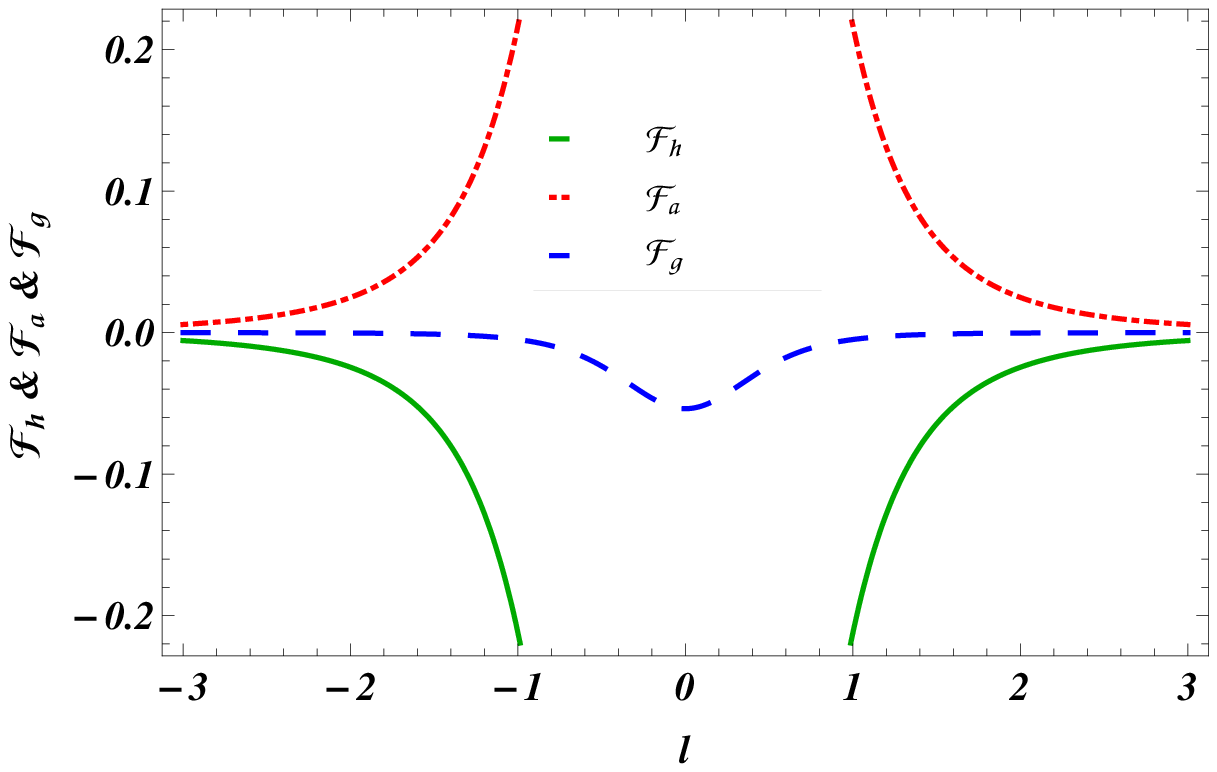, width=.4\linewidth,
height=2.1in}\epsfig{file=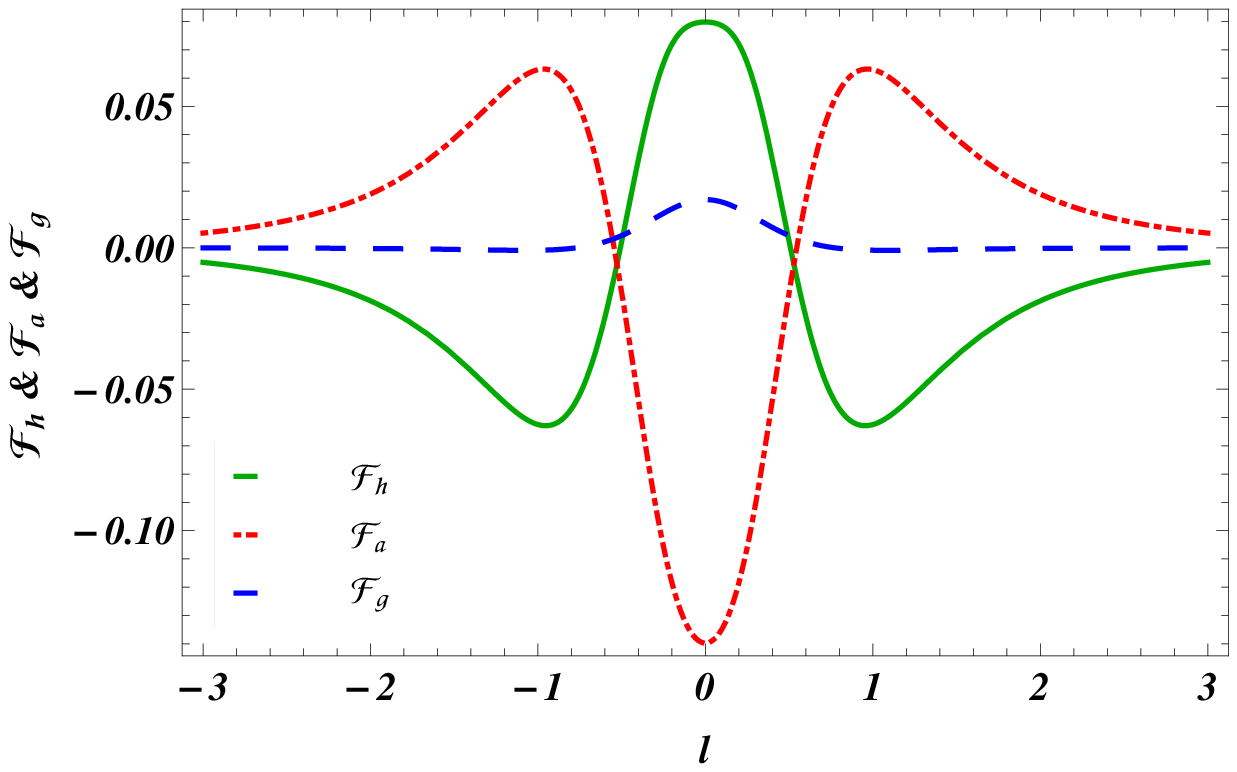, width=.4\linewidth,
height=2.1in}
\centering \epsfig{file=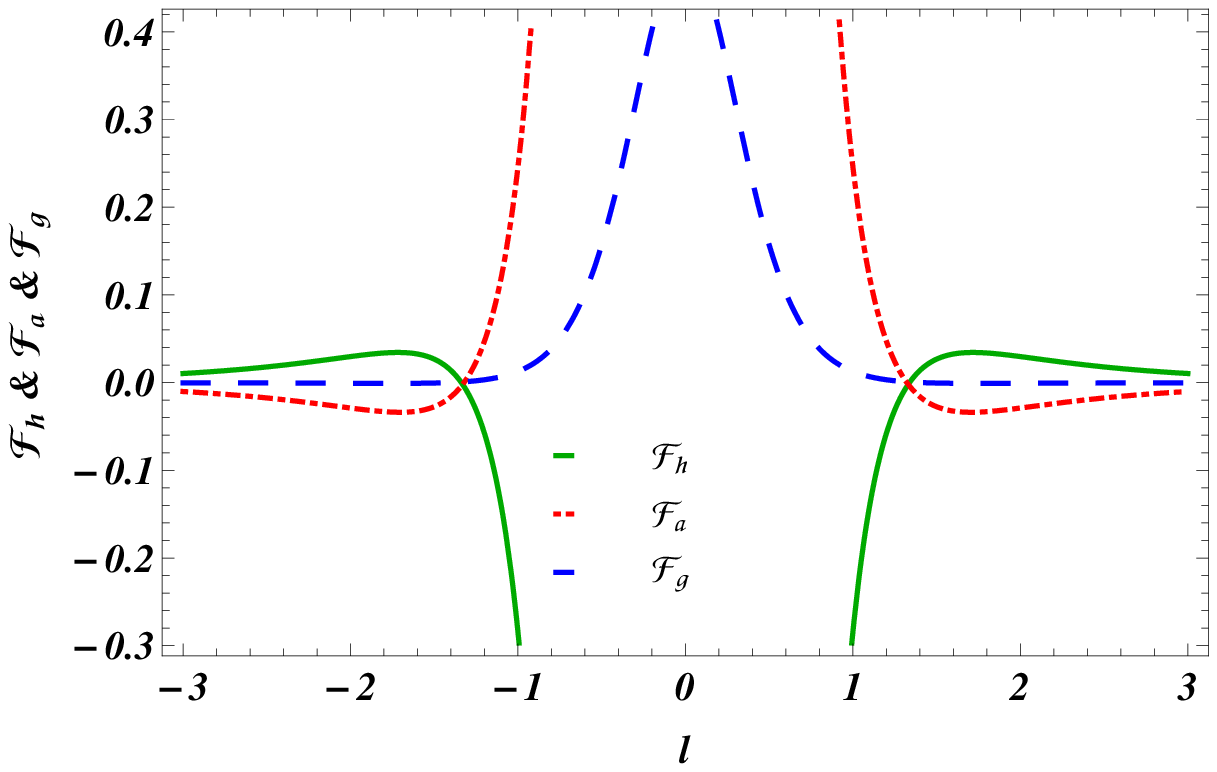, width=.4\linewidth,
height=2.1in}\epsfig{file=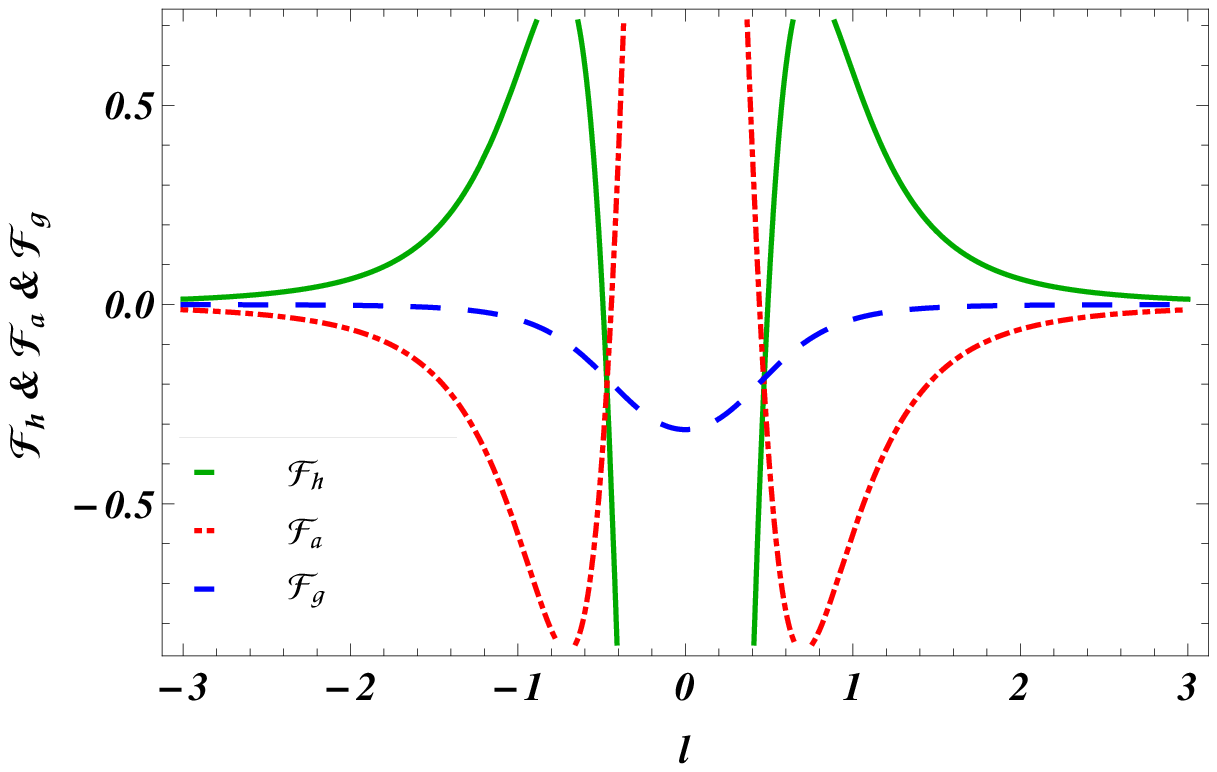, width=.4\linewidth,
height=2.1in} \caption{\label{Fr1} shows the behavior of forces shape function -I (first row) and embedded shape function-II (second row).}
\end{figure}
\section{Equilibrium Analysis via TOV Equation}
The current Section provides the equilibrium analysis for two different dark matter halo models in the background of $f(Q)$ gravity. In order to check the stability, we shall reshape the Tolman-Oppenheimer-Volkoff (TOV) equation for $f(\mathcal{R},\mathcal{G})$ gravity as:
\begin{equation}\label{r1}
\frac{2\Delta}{r}-\frac{dp_{r}}{dr}-\frac{\epsilon'(r)}{2}(\rho +p_r)=0.
\end{equation}
In the above Eq. (\ref{r1}), $\Delta$ denotes the difference of the pressure components, like $p_{t}-p_{r}$. Further, the above TOV equation can be rearranged as
\begin{equation}\label{r2}
\mathcal{F}_g= \frac{\epsilon'(r)}{2}(\rho +p_r),\;\;\;\;\; \mathcal{F}_h= \frac{dp_{r}}{dr},\;\;\;\;\; \mathcal{F}_a= \frac{2\Delta}{r},
\end{equation}
where $\mathcal{F}_g$, $\mathcal{F}_h$ and $\mathcal{F}_a$ represent the anisotropic, hydrostatic and gravitational forces, respectively.
Furthermore, the above Eq. (\ref{r2}) can be reshaped as:
\begin{equation}\label{r3}
\mathcal{F}_g +\mathcal{F}_h+\mathcal{F}_a=0.
\end{equation}

Fig. (\ref{Fr1}) indicates the combined graphical behavior of the forces for both shape function -I and embedded shape function-II, respectively. From both rows, it can be noted that the stability of the system via the TOV equation is obtained within the anisotropic and gravitational forces against the hydrostatic forces for both models. However, initially, instability prevails to some extent, and then gradually forces are seen balanced each other at around the radius of the throat.

\section{Conclusion}

This paper investigates the physical viable WH solutions in $f(\mathcal{R},\mathcal{G})$
gravity in the context two different generic shape functions. For
this purpose, we have considered an anisotropic matter distribution
and computed the matter components for the respective WH solution.
To look into the embedded WH models, we have adopted two different
methodologies, i.e., Karmarkar condition and Ellis-Bronniokv
embedded geometry. The behavior of energy conditions have also been
discussed for these embedded solutions of WH geometry by choosing a
particular model for the red-shift function. We provide a graphical
analysis in order to explore the possibility for existence of WH
solutions as well as their physical behaviors. The results can be
summarized as follows.
\begin{itemize}
  \item It is worth mentioning that the model parameter $\lambda$ has a
significant role in the current scenario as it yields different
possibilities for the WH existence in the range $-1\leq \lambda\leq
1$.
  \item  We have also illustrated the results for energy conditions
graphically. In Fig.(\ref{F1}), we have the energy density plot
through Eq.(\ref{13}) by assuming $-1\leq \lambda\leq 1$ and the
radial distance $-3\leq l\leq3$.

  \item The null energy condition, i.e.,
$\rho+p_{r}$ is presented graphically in Fig.(\ref{F2}) which is
strongly violated with the choices $-1\leq \lambda\leq 1$ and $-\leq
l\leq1$ for the embedded shape function-I (left) and function-II
(right) showing the presence of exotic matter.

  \item Furthermore,
Fig.(\ref{F3}) depicts the violation of $\rho+p_{t}$ in the small
ranges of $\lambda$.

  \item We have provided plots for the violation of
strong energy condition $\rho+p_{r}+2p_{t}$ and trace energy
condition $\rho-p_{r}-2p_{t}$ in Figs.(\ref{F4}) and (\ref{F5}),
respectively.

  \item This violation actually confirms the possibility of WH
existence for embedded shape function -I (left) and embedded shape
function-II (right).

  \item Figs.(\ref{F6}) and (\ref{F7}) show some other
energy conditions like $\rho-\mid p_{r}\mid$ and $\rho-\mid
p_{t}\mid$, respectively.
\end{itemize}
Thus the choice of these generic shape
functions significantly affects the physical analysis and viability
of the embedded WH solutions.
\section*{Conflict Of Interest statement }
The authors declare that they have no known competing financial interests or personal relationships that could have appeared to influence the work reported in this paper.
\section*{Data Availability Statement} This manuscript has no associated data, or the data will not be deposited. (There are no observational data related to this article. The necessary calculations and graphic discussion can be made available on request.)

\section*{Appendix-I}
\begin{eqnarray*}
\psi _1&&=6 r^5 b^{(3)}(r)+r^3 (8 \zeta -109 r) b''(r)+r \left(-12 \zeta ^2
+985 r^2-178 \zeta  r\right)\\&& b'(r)+8 \zeta ^3+1494 r^3-502 \zeta  r^2-52
\zeta ^2 r,\\
\psi _2&&=320 r^5 b^{(3)}(r)-128 r^3 (41 r-3 \zeta ) b''(r)+32 r^2
(367 r-51 \zeta ) \\&&b'(r)^2-32 r b'(r) \left(33 r^3 b''(r)+20 \zeta ^2
-1298 r^2+236 \zeta  r\right)\\&&+128 \zeta ^3+40 \lambda
r^7-139 \zeta  \lambda  r^6-4 \zeta ^2 \lambda  r^5+4 \zeta ^3 \lambda
r^4+18432 r^3\\&&-6912 \zeta  r^2-768 \zeta ^2 r,\\
\psi _3&&=192 \zeta  r b'(r)^4+64 \zeta  (19 r-2 \zeta ) b'(r)^3+4 \lambda
r^2 \left(r^5 b^{(3)}(r) (\zeta +2 r)\right.\\&&\left.+r^3 \left(2 \zeta ^2-4 r^2-5 \zeta
r\right) b''(r)+\zeta ^4-24 \zeta ^2 \times r^2\right)
+r b'(r)^2 \left(-128 \right.\\&&\left.\zeta  r b''(r)-12 \lambda  r^5-8 \zeta  \left(\lambda
r^4-128\right)+5 \zeta ^2 \lambda  r^3\right)+2 r^2 b'(r) \left(b''(r)
\right.\\&&\left.\left(2 \lambda  r^5+\zeta  \left(\lambda  r^4-64\right)\right)
+r \left(-2 \zeta ^3 \lambda +r^5+8 \lambda  r^3+24 \zeta  \lambda  r^2-24
\zeta ^2 \lambda  r\right)\right),\\
\psi _4&&=4 \zeta ^3+931 r^3-287 \zeta  r^2-28 \zeta ^2 r,\\
\psi _5&&=-544 \zeta  r^3 b''(r)-128 \zeta ^3+4 \lambda  r^7+16 \zeta
\lambda  r^6-19 \zeta ^2 \lambda  r^5-2 \zeta ^3 \lambda  r^4\\&&+5888 \zeta
r^2-1024 \zeta ^2 r,
\end{eqnarray*}
\begin{eqnarray*}
\psi _6&&=2 r \left(b^{(3)}(r) \left(2 \lambda  r^6+\zeta  \lambda  r^5+32
\zeta  r\right)+\zeta  \lambda  \left(2 \zeta ^3+12 r^3-60 \zeta  r^2
\right.\right.\\&&\left.\left.-\zeta ^2 r\right)\right)+b''(r) \left(64 \zeta ^2-6 \lambda  r^6-9 \zeta
\lambda  r^5+4 \zeta ^2 \lambda  r^4-896 \zeta  r\right),\\
\psi _7&&=3 r^3 (5 \zeta +r) b''(r)+r \left(46 \zeta ^2-81 r^2-217 \zeta
r\right) b'(r)+4 \zeta ^3-36 r^3\\&&-467 \zeta  r^2+156 \zeta ^2 r,\\
\psi _8&&=-64 r^3 (31 \zeta +2 r) b''(r)+32 r \left(-16 \zeta ^2+70 r^2
+141 \zeta  r\right) b'(r)^2\\&&-64 r b'(r) \left(3 r^2 (2 \zeta +r) b''(r)
+88 \zeta ^2-52 r^2-426 \zeta  r\right)-128 \zeta ^3
\\&&+44 \lambda  r^7-105 \zeta  \lambda  r^6-20 \zeta ^2 \lambda  r^5-4
\zeta ^3 \lambda  r^4+21248 \zeta  r^2-7552 \zeta ^2 r,\\
\psi _9&&=-2 b''(r) \left(6 \lambda  r^8+5 \zeta  \lambda
r^7+\zeta^2 \lambda  r^6+320 \zeta ^2 r^2\right)+32 \zeta  \left(-12
\zeta ^2\right.\\&&\left.+16 r^2+109 \zeta  r\right) b'(r)^2+96 \zeta  r (2 \zeta +r)
b'(r)^3\\&&+b'(r) \left(-64 \zeta  r^2 (5 \zeta +r) b''(r)-1472
\zeta ^3+20 \lambda  r^7+28 \zeta  \lambda  r^6\right.\\&&\left.-25 \zeta ^2 \lambda
r^5-2 \zeta ^3 \lambda  r^4+7552 \zeta ^2 r\right)-896 \zeta
^3+r^9\\&&+2 \zeta  r^8+16 \zeta  \lambda  r^6-80 \zeta ^2 \lambda
r^5-18 \zeta ^3 \lambda  r^4-4 \zeta ^4 \lambda  r^3+2304 \zeta ^2 r
\end{eqnarray*}
\begin{eqnarray*}
\psi _{10}&&=4 r^2 \left(\zeta  r \left(-\zeta ^3 \lambda
+r^5-16 \zeta  \lambda  r^2-4 \zeta ^2 \lambda  r\right)-b''(r)
\left(4 \lambda  r^6\right.\right.\\&&\left.\left.+4 \zeta  \lambda  r^5+\zeta ^2 \left(\lambda
r^4+64\right)\right)\right)+b'(r)^2 \left(-256 \zeta ^3
+12 \lambda  r^7+12 \zeta  \lambda  r^6\right.\\&&\left.-9 \zeta ^2 \lambda  r^5
+2304 \zeta ^2 r\right)+384 \zeta ^2 r b'(r)^3-4 b'(r) \left(b''(r)
\left(2 \lambda  r^8+\zeta  \lambda  r^7\right.\right.\\&&\left.\left.+64 \zeta ^2 r^2\right)
+64 \zeta ^3-8 \lambda  r^7-8 \zeta  \lambda  r^6+8 \zeta ^2 \lambda  r^5
+\zeta ^3 \lambda  r^4-512 \zeta ^2 r\right),\\
\psi _{11}&&=128 \zeta ^4+196 r^5+(524 \zeta +343) r^4-2 \zeta  (96 \zeta +497)
r^3+1512 \zeta ^2 r^2\\&&-792 \zeta ^3 r,\\
\psi _{12}&&=3 r^2 \left(-8 \zeta ^3+4 r^4+(12 \zeta +7) r^3-14 \zeta  r^2
+24 \zeta ^2 r\right) b''(r)-2 \left(16 \zeta ^4\right.\\&&\left.+162 r^5+(250 \zeta +259)
r^4-2 \zeta  (28 \zeta+283) r^3+696 \zeta ^2 r^2-240
\zeta^3 r\right) b'(r) \\&&-2 \left(184 \zeta ^4+170 r^5+(646 \zeta +322) r^4
-\zeta  (244 \zeta +2591) r^3\right.\\&&\left.+2350 \zeta ^2 r^2-1040
\zeta ^3 r\right),
\end{eqnarray*}
\begin{eqnarray*}
\psi _{13}&&=-192 \zeta  r^5 b^{(3)}(r)-16 r^2 \left(-32 \zeta ^3
+10 r^4+(42 \zeta +19) r^3-244 \zeta  r^2\right.\\&&\left.+100 \zeta ^2 r\right) b''(r)
+32 r \left(-14 \zeta ^3+35 r^4+(37\zeta +41)
r^3-2 \zeta  (2 \zeta +55) r^2\right.\\&&\left.+98 \zeta ^2 r\right) b'(r)^2-16 b'(r)
\left(3 r^3 \left(4 \zeta ^2+2 r^3+(2 \zeta +3) r^2-4 \zeta  r\right)
b''(r)\right.\\&&\left.-40 \zeta ^4-266 r^5-2 (257 \zeta +244) r^4+7
\zeta  (16 \zeta +317) r^3-1610 \zeta ^2 r^2\right.\\&&\left.+544 \zeta ^3 r\right)
+2816 \zeta ^4-54 \lambda  r^8+197 \zeta  \lambda  r^7-52
\zeta ^2 \lambda  r^6+4 \zeta ^3 \lambda  r^5+1152 r^5\\&&+8448 \zeta  r^4
+2304 r^4-3264 \zeta ^2 r^3-52160 \zeta  r^3+38208 \zeta ^2 r^2-14400 \zeta ^3 r,\\
\psi _{14}&&=8 \lambda  r^{11} b^{(3)}(r)+4 \zeta  \lambda  r^{10} b^{(3)}(r)
+320 \zeta ^2 r^5 b^{(3)}(r)-2 r^2 b''(r) \left(224 \zeta ^4\right.\\&&\left.+10 \lambda  r^8
+7 \zeta  \lambda  r^7-6 \zeta ^2 \lambda  r^6-32 \zeta  r^4-64
\zeta  (4 \zeta +1) r^3+2688 \zeta ^2 r^2\right.\\&&\left.-704 \zeta ^3 r\right)-16 \zeta r^2
\left(16 \zeta ^2+6 r^3+(6 \zeta -7) r^2-42 \zeta  r\right) b'(r)^3-8
\zeta  r b'(r)^2\\&& \left(3 r^3 (2 \zeta +r) b''(r)-96 \zeta ^3+204 r^4+4
(57 \zeta +85) r^3-2 \zeta  (12 \zeta +433) r^2\right.\\&&\left.+596 \zeta ^2 r\right)+2 b'(r)
\left(16 \zeta  r^3 \left(10 \zeta ^2+5 r^3+
(5 \zeta +9) r^2-25 \zeta  r\right) b''(r)\right.\\&&\left.-256 \zeta ^5+10 \lambda  r^9
+10 \zeta  \lambda  r^8-33 \zeta ^2 \lambda  r^7+6 \zeta ^3 \lambda  r^6
-832 \zeta  r^5-64 \zeta\right. \\&&\times\left. (41 \zeta +26) r^4+128 \zeta ^2
(4 \zeta +143) r^3-9344 \zeta ^3 r^2+3072 \zeta ^4 r\right)\\&&-896 \zeta ^5+r^{11}
+3 \zeta  r^{10}-2 \zeta ^2 r^9+64 \zeta  \lambda  r^8-328 \zeta ^2 \lambda
r^7+92 \zeta ^3 \lambda  r^6-8 \zeta ^4 \lambda  r^5\\&&-2304 \zeta ^2 r^4+896
\zeta ^3 r^3+18432 \zeta ^2 r^3-12672 \zeta ^3 r^2+4096 \zeta ^4 r,\\
\psi _{15}&&=16 \zeta  r^3 (2 \zeta +r) b''(r)-320 \zeta ^4+8 \lambda
r^8+10 \zeta  \lambda  r^7-3 \zeta ^2 \lambda  r^6+512 \zeta  r^4\\&& +128 \zeta
 (5 \zeta +8) r^3-64\zeta ^2 (\zeta +65) r^2+1728 \zeta ^3 r,\\
\psi _{16}&&=2 r^3 b''(r) \left(64 \zeta ^3+2 \lambda  r^7+\zeta  \lambda
r^6+32 \zeta  r^3+32 \zeta  (\zeta +2) r^2-320 \zeta ^2 r\right)\\&& -128 \zeta^5
+r^{11}+\zeta  r^{10}+32 \lambda  r^9+40 \zeta  \lambda  r^8-72 \zeta ^2
\lambda  r^7+12 \zeta ^3 \lambda  r^6\\&&-1024 \zeta ^2 r^4+128 \zeta ^2 (\zeta
+74) r^3-3968 \zeta ^3 r^2+1280 \zeta ^4 r,\\
\psi _{17}&&=r^3 \left(2 b^{(3)}(r) \left(32 \zeta ^2+2 \lambda
r^6+\zeta \lambda  r^5\right)+\zeta  \left(-2 \zeta ^3 \lambda
+r^5-\zeta  r^4\right.\right.\\&&\left.\left.-64 \zeta \lambda  r^2+20 \zeta ^2 \lambda
r\right)\right)-2 b''(r) \left(32 \zeta^4\right.\\&&\left.+6 \lambda
r^8+4 \zeta  \lambda  r^7-3 \zeta ^2 \lambda  r^6-32 \zeta ^2
r^3+448 \zeta ^2 r^2-96 \zeta ^3 r\right)
\end{eqnarray*}

\end{document}